# A Multi-theoretical Kernel-based Approach to Social Network-based Recommendation

Xin Li [a,*], Mengyue Wang [a], and T-P Liang [b]

[a] Department of Information Systems, City University of Hong Kong, Hong Kong

(Xin.Li.PhD@gmail.com, menwang@student.cityu.edu.hk)

[b] Department of Information Management, National Chengchi University, Taipei, Taiwan

(tpliang@faculty.nsysu.edu.tw)

Corresponding Author:

Xin Li, PhD

Assistant Professor

Department of Information Systems

City University of Hong Kong

83 Tat Chee Ave, Kowloon

Hong Kong

Phone: +852-3442-4446

Fax: +852-3442-0370

Email: Xin.Li.PhD@gmail.com

# A Multi-theoretical Kernel-based Approach to Social Network-based Recommendation

**Abstract**: Recommender systems are a critical component of e-commerce websites. The rapid development of online social networking services provides an opportunity to explore social networks together with information used in traditional recommender systems, such as customer demographics, product characteristics, and transactions,. It also provides more applications for recommender systems. To tackle this social network-based recommendation problem, previous studies generally built trust models in light of the social influence theory. This study inspects a spectrum of social network theories to systematically model the multiple facets of a social network and infer user preferences. In order to effectively make use of these heterogonous theories, we take a kernel-based machine learning paradigm, design and select kernels describing individual similarities according to social network theories, and employ a non-linear multiple kernel learning algorithm to combine the kernels into a unified model. This design also enables us to consider multiple theories' interactions in assessing individual behaviors. We evaluate our proposed approach on a real-world movie review dataset. The experiments show that our approach provides more accurate recommendations than trust-based methods and the collaborative filtering approach. Further analysis shows that kernels derived from contagion theory and homophily theory contribute a larger portion of the model.

**Keywords**: Social network, Recommender systems, Non-linear multiple kernel learning

# 1. Introduction

In recent years, there has been a surge in online social networking services. Facebook, Twitter, and LinkedIn have become a part of our social life. Various e-commerce websites also include social networking as a value-added service. These advances not only further engage users and change their behavior, but also enable companies to collect richer information about users for marketing and decision making. Recommender systems have especially evolved with the development of Web 2.0.

Recommender systems are a successful application introduced in the e-commerce era to suggest products, services, and contents to users based on their history with the website. They alleviate the information overload problem, reduce users' efforts in searching products, improve customer satisfaction [32], and promote sales [10]. Traditional recommender systems rely on user characteristics [39], item attributes, and user-item relations [38] to infer user preferences [48]. Since users' activities on e-commerce websites are often infrequent, recommendation algorithms often face the sparsity problem, where transactions are too few as compared with users and items to build an effective model. From an algorithm perspective, dimension reduction methods, such as matrix factorization [37], have been incorporated to alleviate this problem. Another solution to this problem is to incorporate information to enrich user profiles.

Online social networking services provide us with additional social relation information between users, which can help alleviate the sparsity problem [20]. Moreover, they enable new applications in addition to product recommendation. For example, in a social networking website, recommender systems can help identify the advertisements in which a user is interested for advertisement distribution. In mobile applications, where social relations can be inferred from users' communication histories, such as phone calls or SMSs, recommender systems can be used for phone app recommendation. (A related but different problem is to recommend friends to persons in a social network [40], which is essentially a link prediction problem and is not our focus.) In general, it is argued that the existence of social networks enables us to better understand subtle reasons behind users' decisions and to build enhanced decision models [5].

In this paper, we study social network-based recommendation, where we assume users have a social network and the existence of a subset of their rating matrix. The goal is to assess the missing part of the rating matrix. A dominant approach in social network-based recommendation is to develop trust models and estimate user interest based on his/her trusted people's preference [43]. This approach is directed by social influence theory. However, several other social network theories can be used to understand the multiple roles of social relations in decisions. For example, homophily theory [44] is one basis of traditional similarity-based recommendation algorithms. As found by [1], social influence explains less than half of the reasons why people accept their friends' opinions. However, studies on considering other theories are sparse in social network-based recommendation.

In our research, we aim to systematically exploit the information embedded in social networks that can help recommendation based on major social network theories, including contagion theory (social influence theory and regular equivalence theory), collective action theory, and homophily theory (status homophily and value homophily). We propose a multi-theoretical kernel-based approach that enables us to map various theories into a uniform kernel form and convert the recommendation problem to a kernel-based machine learning problem. In this research, we design and select kernels corresponding to major social network theories and then adapt a non-linear multiple kernel learning (NLMKL) method [13] to combine the multiple kernels for recommendation. We conducted experiments on a real-world movie review dataset and found significant performance improvements from this multi-theoretical approach as compared with trust-based approaches. We also found that contagion theory and homophily theory contribute more to recommendation in our proposed approach. In general, our findings support the use of social network information to enhance recommender systems.

Our major contribution in this paper is three-fold. First, we formalize a multi-theoretical approach for social network-based recommendation, which is more theoretically sound and practically effective than trust-based approaches. It provides a basis for future extensions by incorporating relevant theories and selecting kernels that can better fit the theories. Second, we consolidate major social network theories for recommendation, which bridges the parallel explorations in social science and data mining on this problem. Third, we propose an impact distribution kernel to represent regular equivalence theory and show the superiority of this theory and kernel for recommendation in our experimental study.

## 2. Background and Literature Review

### 2.1 Social Network-based Recommendation

#### 2.1.1 Trust Approach

Making recommendations with social network information is an important issue recently. In extant literature, most research along this line was framed under the concept of trust [11, 57, 62]. While trust can be defined upon products, vendors, and sellers, in social network-based recommendation it refers to social trust. So if a person trusts another person, s/he would make decisions similar to that person's [20]. The most reliable trust values are collected from users [21]. Given a set of trustees, one person's rating on a product can be estimated by the average of the trustees' weighted by the trust level [4, 21].

Collecting interpersonal trust is labor intensive. Computational methods were often used to enrich trust based on the underlying social network. The trust between indirectly connected individuals can be inferred based on direct trusts along the social paths that connect them [59]. It can be implemented by conducting an iterative breadth-first search to find one's (direct and indirect) neighbors and aggregate the neighbors' trust values [20]. It can also be implemented through a propagation mechanism [23] with a damping factor controlling distance from neighbors [24] for aggregation. The propagation can go through the entire social network or stop within a number of steps from the starting node [61].

Social trust has also been combined with other features to improve the effectiveness of recommendation. For example, Hess and Schlieder combined user trust values with item quality measures (document visibility) for document recommendation [26]. Guy et al. considered the effect of time on user selection in this problem [24]. Jamali and Ester included trust as user features in a matrix factorization framework for recommendation [31]. Victor et al. also proposed the distrust concept to complement trust in dealing with controversial opinions [58].

**2.1.2 Similarity & Other Approaches**

Attributing all aspects of social networks to trust is arbitrary. In social networks, social influence can appear together with homophily, confounding effects, and simultaneity, among other factors [2].

Table 1. Summary of Previous Social Network-based Recommendation Studies

| Studies | Social Network Info. | | Other Info. | Prediction Model |
|---|---|---|---|---|
| | Type | Manipulation | | |
| Golbeck and Hendler 2006 [21] | Direct trust | User-reported | - | Weighted average |
| Jamali and Ester 2010 [31] | | User-reported | - | Matrix factorization |
| Golbeck 2009 [20] | Indirect trust | Aggregation (breadth-first search) | - | Weighted average |
| Avesani et al. 2005 [4] | | Propagation | - | Weighted average |
| Hess and Schlieder 2008 [26] | | Propagation | Document popularity | Weighted average |
| Walter et al. 2008 [59] | | Propagation | - | Weighted average |
| Yuan et al. 2010 [61] | | Propagation | - | Weighted average |
| Guha et al. 2004 [23] | Trust / distrust | Propagation | - | Voting |
| Victor et al. 2011 [58] | | Propagation | - | - |
| Guy et al. 2009 [24] | Trust + similarity | - | Time | Weighted average |
| Jamali and Ester 2009 [30] | | Propagation | - | Random walk |
| Symeonidis et al. 2011 [56] | Similarity | - | Co-commenting, co-rating | Weighted average |
| Seth and Zhang 2008 [53] | | Clustering | Text | Bayes Network |
| Pham et al. 2011 [49] | | Clustering | Co-authorship | Weighted average |
| He and Chu 2010 [25] | Others | - | User preferences, item popularity | Probabilistic model |
| Yuan et al. 2011 [60] | | - | Membership | Matrix factorization |

Similarity (i.e., homophily) of individuals has often been used in collaborative filtering. The similarity of social networks (adjacency matrix) around a person has also been used together with co-commenting and co-rating relations for recommendation [56]. Jamali and Ester took a random walk approach to combine similarity of users with trust between users for recommendation [30].

After defining user similarity measures, clustering is a common practice to highlight relevant users and improve recommendation effectiveness. Social network-based recommendation also adopts this practice. For example, Pham et al. conducted network-based clustering and used the clusters as neighborhoods in a collaborative filtering paradigm [49]. Seth et al. took a Bayes user decision model in a message recommendation context, in which the cluster structure of social networks is used to characterize known information (message) of a user [53].

Without differentiating the exact factor causing friends' common choices, Yuan et al. minimized the difference between the ratings of connected people in a social network via the regularization model under a matrix factorization framework for top-N recommendation [60]. He et al. also looked at correlations between friends' choices and developed a probabilistic model to aggregate user preferences, item popularity, and social influence for personalization [25].

Table 1 summarized major studies employing social network information. The trust-based approach is the major approach in social network-based recommendation. A few previous studies have explored other aspects of social relations in this problem. Limited studies have systematically exploited the multi-facets of social network information in a unified framework for recommendation.

**2.2 Social Network Theories Relevant to Recommendation**

In addition to trust, there are other theories that can be used to explain behaviors in social networks. We want to systematically explore additional social network theories to enhance our design of recommender systems. In particular, we choose the contagion theory, collective action theory, and homophily theory, since they all explain how people form opinions or take action in a social network. Modeling them in a recommendation task would help us understand how social contexts influence persons' decisions.

**2.2.1 Contagion Theory**

Contagion theory explains how one person's attitude and behavior development is infected based on information, attitudes, and behaviors of others in the network [45]. Contagion theory has two major variants, contagion by cohesion and contagion by structural equivalence [9]. The **social influence theory** [18] is a major example of contagion by cohesion, in which influence is exerted through direct communication. Individuals' thoughts, feelings, attitudes, or behaviors are shifted by their more powerful social contacts. Under this theory, more effort is required to affect those who are farther away in the network [63].

**Regular equivalence theory** [47], as an example of contagion by structural equivalence, inspects whether people have similar patterns in their social relationships (instead of identical contacts) to assess their interference [52]. Pattison argued that people who are regularly equivalent are more likely to have similar social cognitions because "cognitive processes may directly involve the individual's perceptions of his or her social locale" [47]. Individuals with similar social locale are likely to receive similar information, develop similar expectations and biases, and have similar perceptions on subjects. It provides us with an opportunity to look at social networks differently from the existing social trust perspective.

Previous trust-based approaches are generally based on social influence theory rather than regular equivalence theory. Trust-based approaches consider a high influence between people to indicate a high possibility of similar interests. Regular equivalence theory considers similar social context as the reason for similar interests. If two persons have similar impact on a group of people (or are influenced by the same group of people) but do not have a strong impact on each other, they will be considered making similar decisions according to regular equivalence theory but not in the trust-based approach.

**2.2.2 Collective Action Theory**

Contagion theory generally inspects influences between pairs of persons. In social networks, there are scenarios where one person adapts his/her opinions according to a social group. In this case, mutual interest (instead of self-interest) is the objective for people seek to maintain [35] and the output can be quite different from executing individual influence gradually [17]. **Collective action theory** focuses on the possibility of benefits from coordinated action [42]. This theory has been used to study the adoption of innovations in organizations [41, 51].

**2.2.3 Homophily Theory**

Homophily theory [44] studies how the similarity between individuals affects their behavior. Early literature focuses on its role in enabling the formation of social relations [28]. More recently, marketing literature has placed homophily center stage and demonstrated its substantial effect on a recipient's willingness to accept a source's recommendations [14, 19].

Lazarsfeld and Merton (1954) distinguished two types of homophily: **status homophily** and **value homophily**. Status homophily is based on informal, formal, or ascribed status, including society-ascribed demographic characteristics like race, ethnicity, gender, or age, and acquired demographic characteristics like religion, education, occupation, or behavior patterns. Value homophily is based on values, attitudes, and beliefs including the wide variety of internal states presumed to shape our orientation toward future behavior [44], which can be captured from individual's claims.

**2.3 Research Gaps**

From the review, we identify the following research gaps in social network-based recommendation:

1. Social trust (or social influence) is the major approach employed in previous social network-based recommendation research. Studies in other directions are limited. Such a single theoretical lens may limit the design of more effective recommendation algorithms.

**2.** Several other theories provide possibilities of developing useful features to improve recommendation effectiveness. However, how these theories can be modeled and integrated to create a better recommendation method is not straightforward.

In previous research, Arazy et al. also took a theory-testing perspective and investigated the effect of multiple theories in different stages of a decision process to aid recommendation algorithm design [3]. Our paper differs from their in that we focus on an effective recommendation algorithm (together with the understanding of different theories' prediction power under this approach).

## 3. The Proposed Multi-Kernel Approach

In order to fill the aforementioned research gaps, we propose a multi-theoretical approach for social network-based recommendation (Fig. 1). This approach converts the recommendation problem to a kernel-based machine learning problem and employs kernels to capture information reflecting people's interests in light of different social network theories.

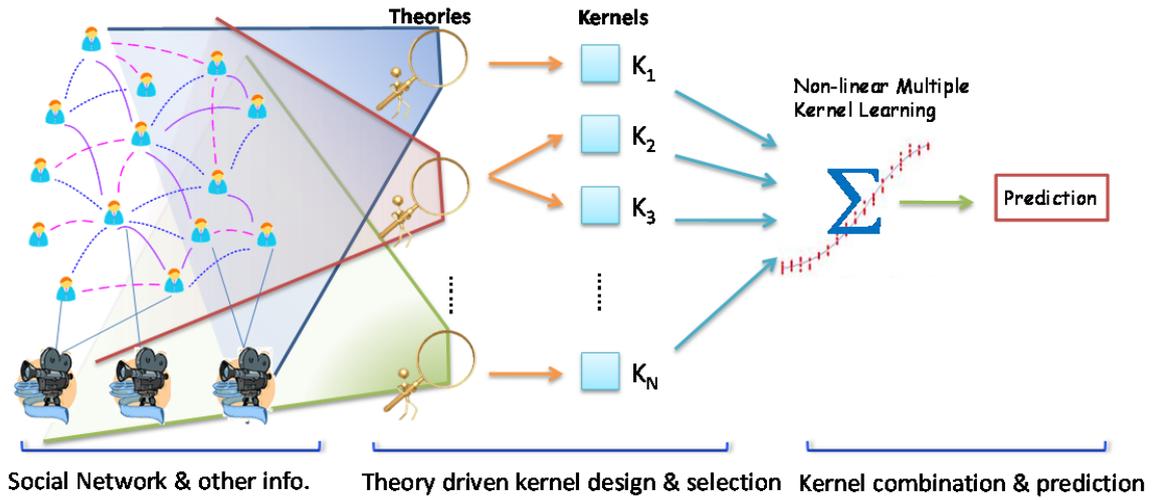

Fig. 1. A Multi-theoretical Approach for Social Network-based Recommendation

The core of the proposed approach is the kernels that reflect different theoretical lens. Hence, the first step in our approach is to select/design kernels according to social network theories. Since social network theories generally describe interactions among users, it is difficult to code these entangled relations as user profile features. By taking the kernel-based paradigm, each user is treated as a data instance and their relations can be captured in a kernel (i.e., a kernel function), which is a similarity measure defined upon these data instances. The similarity measure implicitly maps data to a feature space $H$ (named reproducing kernel Hilbert space, RKHS). So, mathematically, each kernel is equivalent to a type of social network-based feature developed from its corresponding theory.

After specifying kernels, we need to select a kernel machine, i.e., a learning algorithm, to take the kernels as input and build regression/classification models. If we have only one kernel (on users), we can tackle the rating prediction problem using support vector regression (SVR). We can train a model for each item given known user rating data, which then estimates a user's rating (on the same item) based on similar consumers' ratings (i.e., the support vectors). Since we have multiple kernels, building the model becomes another issue. We need to combine these kernels. We take advantage of the advance in multiple kernel learning and employ the non-linear multiple kernel learning algorithm to optimize the combination of kernels. By doing so, we combine multiple theories while considering their possible interactions. A SVR model is then applied on this combined kernel to predict user ratings and make recommendations.

In this section, we elaborate these two major steps of our approach.

**3.1 Theory-Driven Kernel Design and Selection**

In general, several different types of kernels can be designed for each theory. In this subsection, we give preference to established, intuitive, and effective kernels.

**3.1.1 The Impact Distribution Kernel**

The first kernel we design is an impact distribution kernel that aligns with the regular equivalence theory, in that people having similar patterns of relations tend to favor similar products. The rationale of this kernel is illustrated in Fig. 2. Instead of directly measuring complicated structural patterns of people's social relations, we inspect the pattern of impacts from each person to all other persons in the network. We calculate interpersonal impact based on the network structure. Thus, the distribution of impacts abbreviates structural patterns of people's social relations (including direct and indirect relations) to a certain extent. After capturing this information in a kernel, a kernel machine will be able to estimate individuals with similar distribution of social impacts as having similar interests.

To define the kernel, we conduct random walks with restart on the user social network (say, friendship network) from an arbitrary user $u_i$. The random walks can transmit to $u_i$'s neighborhood with a transition probability. Such transition happens iteratively to farther nodes, while at each step there is a probability for the random walk to return to the original node $u_i$. Eventually, we can get a stationary transition probability $r_{i,j}$ that reflects how much $u_i$'s opinions can reach user $u_j$. In a unidirectional network investigated by this paper, it also reflects $u_j$'s opinions' impact on $u_i$. The distribution of $r_{i,j}$ on all users in the social network reveals the pattern of people's impact structure. As shown in Fig. 2, we then define a kernel from the stationary transition probabilities. (Note that random walk is a popular network modeling approach. It has been used in a variety of ways in previous recommendation studies, such as trust propagation [33], combining trust with other information [7, 29, 30], calculating distances between nodes [16], etc. Some studies employed random walk in a similar manner [46] to derive the influence between objects. However, here we focus on how social structure reflected by social information can be captured in a kernel form and convert this asymmetric measure to a kernel matrix to match the projection of regular equivalence theory.)

Mathematically, if the single-step transition probability matrix is annotated as $\tilde{P}^T$, we initialize the matrix as $p_{ij} = P(s(t+1) = j | s(t) = i) = a_{ij} / \sum_{j=1}^{n} a_{ij}$, where $a_{ij}$ is elements of the social network's adjacency matrix indicating whether $u_i$ and $u_j$ have a relationship (0 for no relationship and 1 for a relationship). For a random walk starting from $u_i$ at the $t$ step, its probability to reach each node can be recorded as a vector $r_i(t) = \alpha \tilde{P}^T r_i(t-1) + (1-\alpha) e_i$   $r_i(0) = e_i$, where $e_i$ is the unit vector and $\alpha$ is the damping factor. By solving this equation, we get: $r_i(t) = (\sum_{k=0}^{t} (\alpha \tilde{P}^T)^k - \alpha \sum_{k=0}^{t-1} (\alpha \tilde{P}^T)^k) e_i$. Since $\lim_{t \to \infty} \sum_{k=0}^{t} (\alpha \tilde{P}^T)^k = (I - \alpha \tilde{P}^T)^{-1}$, the stationary transition probability vector is: $r_i = (1-\alpha)(I - \alpha \tilde{P}^T)^{-1} e_i$, i.e., the stationary transition metric is $R = \{r_{ik}\} = (1-\alpha)(I - \alpha \tilde{P}^T)^{-1}$. Using $r_i$ to represent the probability for random walks reaching other nodes from user $u_i$, we can define a valid impact distribution kernel: $K_{ID} = R^T R$.

In previous literature, various social network centrality measures, such as degree, closeness, and betweenness, were also used to measure people's position and roles in a social network. They are found statistically significant in explaining social phenomena (such as work performance) after controlling for other factors. However, they mainly capture relative social position, which is different from the social structure discussed here. Through small-scale experiments, we find such aggregative social position information are not effective for the recommendation task and do not use them to build kernels in this study.

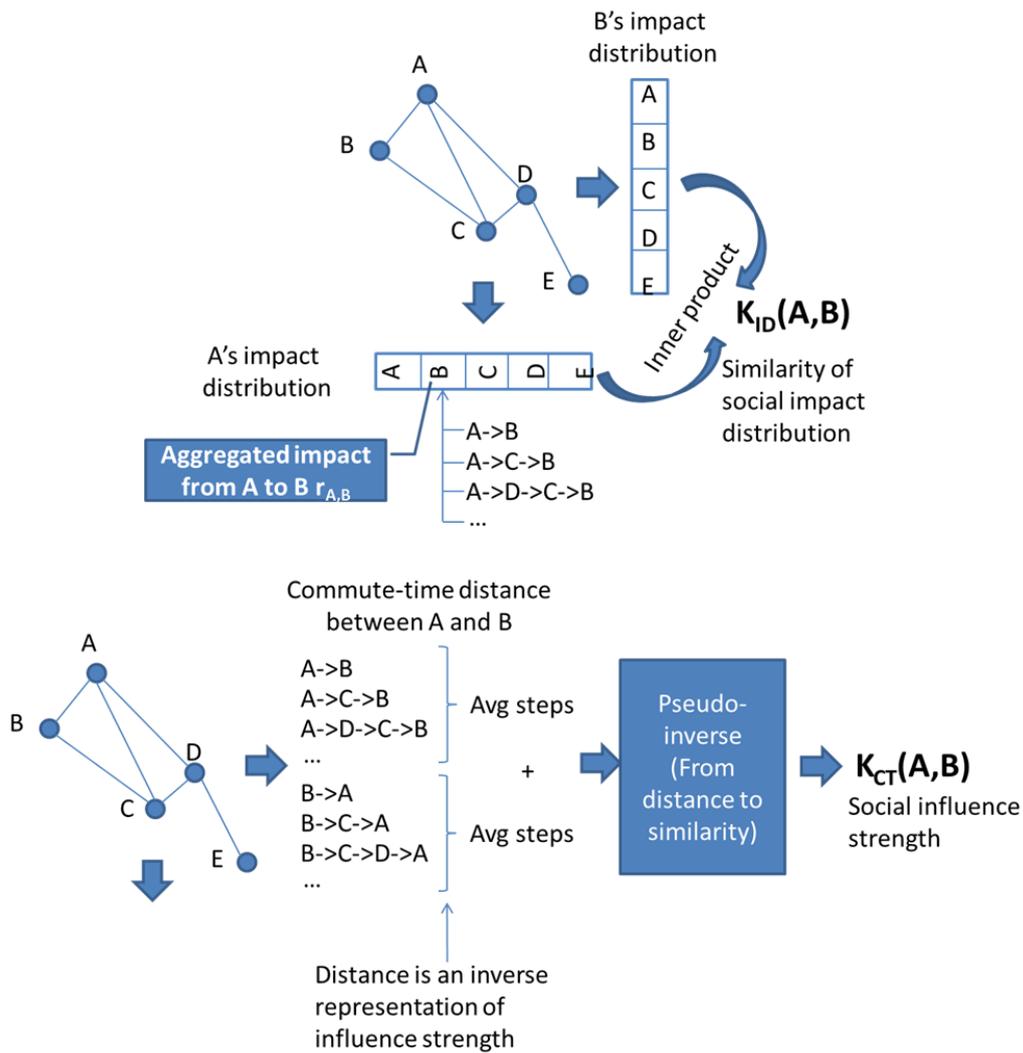

Fig. 2. Comparison of Impact Distribution Kernel and Commute Time Kernel

### 3.1.2 The Commute Time Kernel

The second kernel we adopt is the commute time kernel capturing nodes' topological closeness on a network [16]. Since we do not assume any trust values collected from the people in the social network, the distance between people is a (inverse) representation of interpersonal influence, where shorter distances indicate stronger influences. Capturing such social closeness in a kernel, a kernel machine will estimate that closer individuals on the network have similar interests. In essence, this kernel reflects the social influence theory, which argues that individuals would influence each other on thoughts, and it is easier for closer individuals on a social network to influence each other and form common interests.

Compared with the above impact distribution kernel, which inspects the distribution of one person's impact on all other persons, the commute-time kernel focuses on the influence between a pair of persons and directly converts that influence to a kernel representation (Fig. 2).

As illustrated in Fig. 2, the kernel calculates distance between individuals (i.e., nodes) on a (social) network based on average commute time for random walks to travel from one node to reach another node, and then return to the original one. Commute time is an asymmetric measure that accounts for not only the length of paths between nodes but also the probability for one node taking such a path to influence another node. Since distance metrics cannot be used in kernel-based methods, it is necessary to transform to the RKHS. Thus, the kernel can be obtained by taking a Moore-Penrose pseudo-inverse of the Laplacian matrix (*D-A*) of the graph: $K_{CT} = (D - A)^+ = L^+$, where *A* is the social network's adjacency matrix, $d_{i,i} = [D]_{i,i} = \sum_{j=1}^{n} a_{i,j}$, and $a_{i,j}$ are the elements of *A* indicating relationship between $u_i$ and $u_j$ (0 for no relationship and 1 for a relationship).

Existing trust propagation approaches can also infer trust based on social network topology [23, 24], if there is no prior knowledge of trust. Such measures can be captured in the von Neumann kernel (with a constant damping factor) [34] or the diffusion kernel (with an increasing damping factor) [36] and included in our framework. However, they account for (asymmetric) directional distance from one node to another rather than bidirectional social closeness as in the commute-time kernel (which is implemented by adding the directional distances of two directions). As a result, they do not fit with the bidirectional friendship network setting as well as the commute-time kernel(according to our small-scale experiments). Thus, we do not use them to represent social influence theory in this paper.

**3.1.3 The Community Kernel**

The third kernel we derive is the community kernel. The kernel is grounded on collective action theory, which argues that people make coordinated actions for mutual interests. In the context of recommender systems, a group of people may coordinate their thinking and activities to support or boycott certain items. Here we cluster individuals in communities or social groups based on the topology of their social network, where individuals with dense social relations (and relatively sparse relations to other people) are considered as common interest groups. (It is necessary to cluster people using social network topology to represent collective action theory, since collective action should be established based on people's social interactions. The clustering of people based on other features can also be incorporated in recommendation, which is explained by the homophily theory, as elaborated later.) Such network topology-based clustering has been well studied in the complex network literature. Based on this kernel, a kernel machine will predict individuals in a social group to have similar interests.

We adopt an efficient and effective heuristic method [6] that optimizes network modularity by tweaking the permutations of community allocations of individuals. After identifying the community structure of the network, we define the kernel value of the community kernel as: $k_{COM}(u_i,u_j)=1$ if $u_i$ and $u_j$ belong to the same community. Our design of the community kernel is relatively simple. There are ways to improve it, such as incorporating timestamps into the model and tracing revisions of opinions in a group. We leave that to future research because our main purpose here is to show the feasibility of our approach.

**3.1.4 The Demographic Kernel, Claim Kernel, and Action Kernels**

We design four kernels based on the homophily theory. In general, the homophily theory argues people with similar background have similar preferences [44]. Although this information does not necessarily relate to social networks reported by users, it might reflect hidden social interaction (intentions). Nevertheless, the homophily rationale has been widely used in user-based collaborative filtering and is included in our paper.

We first build a demographic kernel to capture objective attributes of users, including gender, location, education, etc. In general, we apply an inner product function [54] onto these attributes to convert them to the kernel form. For most categorical variables that can be represented as dummy variables, the kernel function is $k_{DEM}(u_i, u_j) = \sum_{k=1}^{n} |c_{ik} \cap c_{jk}| / \sqrt{|c_i||c_j|}$ where $c_i$ and $c_j$ represent attributes for user $u_i$ and $u_j$ respectively. This kernel can be explained using the status homophily theory; here the demographic information represents people's status and can be related to their interests. It is possible to apply a clustering algorithm on such demographic features and group individuals together. After such dimension reduction, each group can be considered as a special type of features to build a kernel. In such a setting, people in a group have similar characteristics and higher probability of common interests.

We then build a claim kernel $K_{CLA}$ based on user-claimed interests, such as favorite product category, favorite characteristics of products, etc. Such claimed interest is often used in a filter mechanism in existing recommender systems. For example, if one is interested in comedic movies, more comedies can be recommended. We also apply an inner product function onto such categorical features to represent them in a kernel form.

We create two action kernels to capture user preference as reflected by their actions. The first one is based on items they rate or purchase. We apply an inner product function to transfer people's shared rating/purchase items to a kernel $k_{ACT1}(u_i, u_j) = |v_i \cap v_j| / \sqrt{|v_i||v_j|}$ ($v_i$ and $v_j$ represent the rated/purchased items for user $u_i$ and $u_j$ respectively). This is essentially the key instrument used in existing collaborative filtering methods. This kernel is equivalent to the Jaccard similarity measure giving the binary input variables of $v$. Pearson's correlation cannot be directly used here, since it does not meet the requirements of being a valid kernel. The second one captures people's habits or bias in their ratings. This reflects whether two users have the same rating scale whether or not they rank favorite products similarly. Here we use the mean of each person's previous ratings as an indicator of their bias. We apply a RBF function [54] on this feature to make it a kernel: $k_{ACT2}(u_i, u_j) = \exp(-\|m_i - m_j\|^2 / 2\sigma^2)$, where $m_i$ and $m_j$ represent mean ratings for the two users $u_i$ and $u_j$ and $\sigma$ is a parameter to be tuned.

Both the claim kernel and the two action kernels have their theoretical basis in the value homophily theory. Different from the status homophily theory, value homophily theory focuses on the impact of subjective claims, knowledge, and perceptions of individuals on the formation of social networks and implication of common interests. Thus, these kernels provide direct assessment of people's interests.

Table 2 summarizes major types of information captured by our selected/designed kernels and their corresponding theories. The first three are mainly based on reported networks; the others are based on different information but may be related to hidden social networks. When implementing these kernels, we make an effort to use most available information in the dataset to increase the power of the model.

Table 2. Mapping Between Kernels and Theories

|  | Name | Captured Information | Theory |
| --- | --- | --- | --- |
| $K_{ID}$ | Impact distribution model | The distribution of people's impact on other persons | Regular equivalence theory |
| $K_{CT}$ | Commute time kernel | Distance between people in social network | Social influence theory |
| $K_{COM}$ | Community kernel | Social groups from social network structure | Collective action theory |
| $K_{DEM}$ | Demographic kernel | Demographic similarities between people | Status homophily theory |
| $K_{CLA}$ | Claim kernel | Self-claimed interest items, categories, etc. | Value homophily theory |
| $K_{ACT1}$ | Action kernel 1 | Action-confirmed interest items, categories, etc. | Value homophily theory |
| $K_{ACT2}$ | Action kernel 2 | People's habit (mean value) on ratings | Value homophily theory |

**3.2 Information Aggregation using Non-linear Multiple Kernel Learning**

**3.2.1 NLMKL Prediction Process**

Suppose we have a kernel $K = \{k(u_i, u_j)\}$ capturing similarities between users, and $\mathbf{y} = [y_1, \ldots, y_N]^T \in R^N$ denotes the vector of training values, which are product ratings. The recommendation task of predicting a user's rating on an item is essentially a regression problem. SVR is an effective sparse kernel machine for this task. SVR aims to find a function *f(u)* to predict the observed value *y(u)* on all training data, while tolerating errors less than a predefined insensitive error margin ε. After minimizing the error on training data, the prediction function will be:

$$y(u) = \sum_{n=1}^{N} (a_n^+ - a_n^-) k(u, u_n) + b$$

where $a_n^+$, $a_n^-$, and $b$ are estimated from training data ($K$ and $\mathbf{y}$) and $a_n^+$ and $a_n^-$ are non-zero only on a small number of users (i.e., support vectors) [55]. So, we can predict any other user's rating as long as we know their kernel value with the support vectors, i.e., $k(u_x, u_n)$.

However, since we capture the different facets of social networks in multiple kernels, we need to combine them into one matrix before they can be used in a SVR framework. An arbitrary approach is to simply add the kernels together to create a composite kernel. A more systematic way is to tune the weight of kernels using the training data when adding kernels together. If the multiple kernels are combined through weighted average, the approach is called linear multiple kernel learning, which is equivalent to directly merging feature spaces in a feature-based method. To further enlarge the dimension of the RKHS (i.e., enlarge the feature space), non-linear MKL [13] were invented to use non-linear functions, namely, multiplication, power, and exponentiation, to create interactions between kernels before combining them.

In this research, we choose to combine kernels using the non-linear MKL (NLMKL) method. Assuming we have p kernels, we first assign a weight η to each kernel, then conduct Hadamard (pointwise) products between all the kernels $d$ times, and finally sum all permutations of kernel products to create a new kernel. The combined kernel is:

$$k_\eta(u_i, u_j) = \sum_{q \in \Re} \eta_1^{q_1} k_1(u_i, u_j)^{q_1} \eta_2^{q_2} k_2(u_i, u_j)^{q_2} \cdots \eta_P^{q_P} k_P(u_i, u_j)^{q_P}$$

where $\Re = \left\{ \mathbf{q} : \mathbf{q} \in Z_+^P, \sum_{m=1}^P q_m = d \right\}$ and $\eta \in R^P$.

In this research, we specify $d=2$ to limit model complexity. Thus, the composite kernel can be written as:

$$k_\eta(u_i, u_j) = \sum_{m=1}^P \sum_{h=1}^P \eta_m \eta_h k_m(u_i, u_j) k_h(u_i, u_j) .$$

Represented in a kernel matrix form, that will be:

$$K_\eta = \sum_{m,h=1}^P \eta_m \eta_h K_m \circ K_h ,$$

where ° is the Hadamard product. As we can see, this combined kernel is a linear combination of multiple components, where each component is the multiplication of some candidate kernels. Since applying the multiplication and summation operations on a valid kernel will generate validated kernels, this new kernel is a valid kernel. The multiplication operation increases the dimensionality of the RKHS. The weight of each group of kernel products can be specified separately. Here, associating a weight with each kernel before conducting multiplication reduces the number of parameters to be tuned later. In order to generate kernel products with an order less than $d$, we include a kernel with all-one elements into this NLMKL approach. For example, when combining P kernels to the order of two, if we include an all-one matrix $K_0$, the final kernel will be:

$$K_\eta = \eta_0 \eta_0 K_0 \circ K_0 + 2\sum_{h=1}^{P} \eta_0 \eta_h K_0 \circ K_h + \sum_{m,h=1}^{P} \eta_m \eta_h K_m \circ K_h$$

$$= \eta_0 \eta_0 + 2\sum_{h=1}^{P} \eta_0 \eta_h K_h + \sum_{m,h=1}^{P} \eta_m \eta_h K_m \circ K_h$$

which contains both the first order of the p kernels and second-order multiplications of $K_h$.

If we have specified the parameters of $\eta$, we can easily calculate $K_\eta$, and use $K_\eta$ in a SVR model to generate rating predictions as aforementioned.

**3.2.2 Kernel Combination Parameter Tuning**

To make the best use of this new combined kernel, we need to tune $\eta$. In this paper, we adopt Cortes et al.'s projection-based gradient descent approach in determining $\eta$ [13]. However, we extend their base learner from Kernel Ridge Regression (KRR) to SVR to better fit our recommendation task.

Given kernel $K_\eta$, the error minimization problem of SVR can be converted to its dual optimization problem by introducing the Lagrange multiplier $\alpha^+, \alpha^- \in R_+^N$:

$$\max_{\alpha^+, \alpha^- \in R_+^N} (\alpha^+ - \alpha^-)^T y - \varepsilon e^T (\alpha^+ + \alpha^-) - \frac{1}{2}(\alpha^+ - \alpha^-)^T K_\eta (\alpha^+ - \alpha^-)$$

subject to $\mathbf{e}^T (\alpha^+ - \alpha^-) = 0$,

where $\alpha^+$ and $\alpha^-$ are the coefficients to be learned from the data, $\mathbf{e}^T$ is an all-one vector, and $\varepsilon$ is the insensitive error margin. If the purpose is to find the most appropriate $K_\eta$, one needs to address a min-max optimization problem [12]:

$$\min_{\eta \in M} \max_{\alpha^+, \alpha^- \in R_+^N} \left(\alpha^+ - \alpha^-\right)^T \mathbf{y} - \varepsilon\, \mathbf{e}^T \left(\alpha^+ + \alpha^-\right) - \frac{1}{2}\left(\alpha^+ - \alpha^-\right)^T \mathbf{K}_\eta \left(\alpha^+ - \alpha^-\right),$$

where $K_\eta$ is defined based on $\eta$ taken from a convex, bounded, and positive set M for efficiency concerns. M is often defined in a $l_2$-norm M=$\{\eta: \eta \in R_+^P, \|\eta - \eta_0\|_2 \leq \Lambda\}$ with parameters $\eta_0$ and $\Lambda$. This min-max optimization problem can be solved in an iterative manner. In each iteration, we can solve the inside maximization problem on a kernel matrix $K_{\eta^*}$ specified by parameter $\eta^*$. The optimal result is:

$$F(\eta^*) = \left(\alpha^{+*} - \alpha^{-*}\right)^T \mathbf{y} - \varepsilon\, \mathbf{e}^T \left(\alpha^{+*} + \alpha^{-*}\right) - \frac{1}{2}\left(\alpha^{+*} - \alpha^{-*}\right)^T \mathbf{K}_\eta \left(\alpha^{+*} - \alpha^{-*}\right),$$

where $\alpha^{+*}$ and $\alpha^{-*}$ are the optimal coefficients given $\eta^*$. Then for the outside minimization problem, $\eta$ is updated by following the gradient descent rule, $\eta = \eta^* - \gamma \nabla F(\eta^*)$. For the second-order non-linear kernel combination used in this paper, we have:

$$\frac{\partial F}{\partial \eta_k} = \partial \left[\left(\alpha^{+*} - \alpha^{-*}\right)^T \mathbf{y} - \varepsilon\, \mathbf{e}^T \left(\alpha^{+*} + \alpha^{-*}\right) - \frac{1}{2}\left(\alpha^{+*} - \alpha^{-*}\right)^T \mathbf{K}_\eta \left(\alpha^{+*} - \alpha^{-*}\right)\right] / \partial \eta_k$$

$$= -\frac{1}{2} \partial \left[\left(\alpha^{+*} - \alpha^{-*}\right)^T \mathbf{K}_\eta \left(\alpha^{+*} - \alpha^{-*}\right)\right] / \partial \eta_k$$

$$= -\frac{1}{2} \partial \left[\left(\alpha^{+*} - \alpha^{-*}\right)^T \left(\sum_{k,h=1}^{P} \eta_k \eta_h K_k \circ K_h\right)\left(\alpha^{+*} - \alpha^{-*}\right)\right] / \partial \eta_k$$

$$= -\frac{1}{2}\left(\alpha^{+*} - \alpha^{-*}\right)^T \left(2\sum_{h=1}^{P} \eta_h K_h \circ K_k\right)\left(\alpha^{+*} - \alpha^{-*}\right)$$

$$= -\left(\alpha^{+*} - \alpha^{-*}\right)^T \left(\sum_{h=1}^{P} \eta_h K_h \circ K_k\right)\left(\alpha^{+*} - \alpha^{-*}\right).$$

We can easily derive this gradient descent function for a combined kernel with any order $d$ as:

$$\frac{\partial F}{\partial \eta_k} = -\frac{1}{2}\left(\alpha^{+*} - \alpha^{-*}\right)^T \left(\sum_{\mathbf{q} \in \mathfrak{R},\ q_k > 0} \eta_1^{q_1} K_1 \circ \cdots \circ q_k \eta_k^{q_k - 1} K_k \circ \cdots \circ \eta_P^{q_P} K_P\right)\left(\alpha^{+*} - \alpha^{-*}\right).$$

If the gradient descent process makes η outside of the bound of M, it will be projected back to the border of M [13]. After iterations of inside maximum and outside minimum, the process converges to a set of $α^+$, $α^-$, and η that together reach the best fitting of **y**.

### 3.2.3 Computational Complexity

The computational complexity of this algorithm can be divided into training and testing stages. Since it adopts the gradient-descent scheme to iteratively update the coefficients of kernels and learn the target classifier, the training complexity depends on these two factors [22]. In each iteration, the cost is essentially for SVR training, which can be $O(n^{2.3})$ with the SMO algorithm [50], where n is the number of training data instances. So for the entire algorithm, the training cost is $O(T*n^{2.3})$, where $T$ is the number of iterations. In this research, non-linear MKL usually converged after no more than 10 iterations.

For the testing process with computed coefficients, these models are general linear models. The computational complexity is proportional to the number of support vectors learned from the training stage, which is generally small in our sparse kernel machines.

## 4. Empirical Evaluation

### 4.1 Dataset

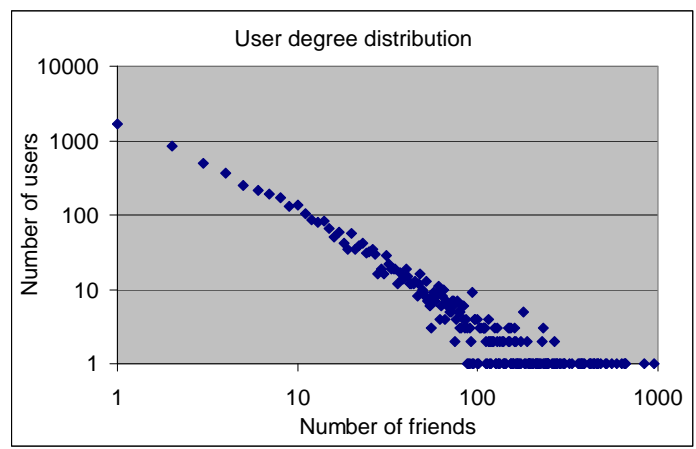

Fig. 3. Distribution of Users and Number of Friends

To evaluate our approach, in November 2010 we collected actual data from www.mtime.com, one of the largest movie review websites in China. The website allows users to rate and comment on movies on a 1 to 10 scale and maintains friendship relationships among users. The website records 5,232 movies released in 2009, which were reviewed by 29,527 users at the time we collected data. Both the number of reviewers of each movie and the number of friends of each user (as shown in Fig. 3) follow a power law distribution, as many social network datasets do. For our study, we chose 452 movies with more than 100 reviews and randomly selected 6,155 reviewers (about 25%) with at least one friend (i.e., connected in the social network). This results in a dataset containing 43,579 inter-user friendships in the network. On average, each person has 14.16 friends (std. dev. 11.50) and rates 5.59 movies (std. dev. 11.16). The average rating is 7.31 (std. dev 1.72).

**4.2 Evaluation Metrics**

To evaluate the performance of our method in predicting consumer ratings, we adopted the widely used Root Mean Square Error (RMSE) measure on the difference between predicted ratings and real ratings:

$RMSE = \sqrt{\frac{1}{m \times n} \sum_{i,j}^{m,n} (P_{i,j} - R_{i,j})^2}$, where $P_{i,j}$ is the predicted rating for user $u_i$ on movie $m_j$ and $R_{i,j}$ is the real rating. A smaller RMSE indicates better performance of a model.

**4.3 Baseline Methods**

To show the superiority of our methods, we chose two traditional trust-based methods as baseline methods for comparison.

*Baseline 1: Neighbor influence model (NI):* As a classic trust-based method, one person's opinion can be influenced by his/her friends. Since our social network does not have any inherent trust-level information, we predict a user's rating of a movie based on the average of his/her friends' ratings:

$P_{v,w} = \frac{1}{|N(v)|} \sum_{u \in N(v)} R_{u,w}$, where $N(v)$ are the friends of user $v$; $P_{v,w}$ is the predicted rating of user $v$ on movie $w$, and $R_{u,w}$ is the rating of user $u$ on movie $w$.

*Baseline 2: Multi-level neighbor influence model (MNI):* As a natural extension of the neighbor influence model, we can include the influence from indirect friends and include multiple levels of neighbors. In this research, we do not have the user-provided trust value on paths between indirect friends. Thus, we account for the decrease of indirect friendship influence power with a constant damping factor α between 0 and 1, and give farther relationships lower weights. The damping factor controls the effect of different levels of neighbors. A larger damping factor makes weight decrease faster, which is equivalent to using a smaller scope of neighbors. In this research, we tune the damping factor to obtain the best performance of the prediction. Eventually we have $P_{v,w} = \sum_{i=1}^{k} \alpha^i R_i$, where $R_i$ is the average ratings of all the users $N_i(v)$ whose shortest path from focal user $v$ is $i$ steps: $R_i = \frac{1}{|N_i(v)|} \sum_{u \in N_i(v)} R_{u,w}$.

Furthermore, we compare our proposed kernel-based approach with a simple user-based collaborative filtering approach [27]. In this approach, we dim the kernels as user similarity functions and get one user's rating of an item based on the average of other people's rating of the item weighted by their similarity to the user as in [15]: $P_{v,w} = \frac{\sum_{u \in V} Sim(v,u) \times R_{u,w}}{\sum_{u \in V} |Sim(v,u)|}$. We conduct this model on the aforementioned individual kernels, which are annotated as CF-S$_{ID}$, CF-S$_{CT}$, CF-S$_{CO}$, CF-S$_{DEM}$, CF-S$_{CLA}$, CF-S$_{ACT1}$, and CF-S$_{ACT2}$, respectively. We also apply it on the average of all kernels, which is named CF-S$_{AVG}$, and the average of most effective individual kernels, including CF-S$_{ID}$-S$_{ACT2}$, CF-S$_{CT}$-S$_{ACT2}$. Some previous studies also include a user bias in the CF method [8]. In our experiments, this design will impair predict performance. We thus keep it in the most basic form to highlight each kernel's effect in this framework. In previous studies, Pearson correlation is often used as the similarity function. However, it is not a valid kernel and cannot be used in kernel-based methods. Nevertheless, for comparison purpose, we include one classic design using Pearson's correlation with user bias, UCF-S$_{Pearson}$, and the combination of most effective kernels and Pearson's correlation with user bias, i.e., UCF-S$_{ID}$-S$_{Pearson}$, and UCF-S$_{ID}$-S$_{Pearson}$.

**4.4 Experimental Procedure**

We randomly split users into 10 groups and conducted 10-fold cross validation in our experiment. The predictions were done one by one for each movie. For the (multi-level) neighbor influence and collaborative filtering models, the rating could be directly predicted. For kernel-based methods, we used

SVR to build models for each movie and estimated the rating of users in the testing dataset (for each fold). We tune the parameters for each kernel for their best performance in individual kernel-based prediction. We then tune parameters for the overall combination in NLMKL as specified above. The predicted ratings were compared with the actual ones to calculate RMSE. To conduct statistical tests, we repeated this process 10 times, which provided us with 10 RMSEs for each method. We then conducted pairwise t tests to compare our proposed approach with the baseline methods.

## 5. Results

Table 3. Comparison of Predictive Performance (10 times' average of RMSE)

| | *Mean* | *Std. Dev.* | | *Mean* | *Std. Dev.* |
|---|---|---|---|---|---|
| **Multi-theoretical Kernel-based Approach** | | | **Collaborative Filtering** | | |
| $K_{ID}$ | 1.396 | 0.001 | CF-$S_{ID}$ | 1.475 | 0.037 |
| $K_{CT}$ | 4.817 | 0.017 | CF-$S_{CT}$ | 2.188 | 0.022 |
| $K_{COM}$ | 6.476 | 0.079 | CF-$S_{COM}$ | 5.632 | 0.095 |
| $K_{DEM}$ | 6.484 | 0.042 | CF-$S_{DEM}$ | 5.835 | 0.100 |
| $K_{CLA}$ | 6.372 | 0.027 | CF-$S_{CLA}$ | 6.007 | 0.077 |
| $K_{ACT1}$ | 1.396 | 0.001 | CF-$S_{ACT1}$ | 5.114 | 0.089 |
| $K_{ACT2}$ | 1.328 | 0.006 | CF-$S_{ACT2}$ | 1.452 | 0.001 |
| Combined Kernel | 1.239 | 0.001 | CF-$S_{AVG}$ | 1.359 | 0.001 |
| | | | CF-$S_{ID}$-$S_{ACT2}$ | 1.275 | 0.001 |
| | | | CF-$S_{CT}$-$S_{ACT2}$ | 1.337 | 0.003 |
| **Trust Methods** | | | **Collaborative Filtering with User Bias** | | |
| NI | 6.757 | 0.018 | UCF-$S_{Pearson}$ | 1.758 | 0.001 |
| MNI | 2.911 | 0.026 | UCF-$S_{ID}$-$S_{Pearson}$ | 1.676 | 0.002 |
| | | | UCF-$S_{CT}$-$S_{Pearson}$ | 1.861 | 0.003 |

Table 3 reports the predictive accuracy of various models. On a 1 to 10 scale, our proposed NLMKL approach achieves a RMSE of about 1.239, which is significantly better than all other models at a 99% confidence interval in pair-wise t test. The results indicate that the multi-kernel based model significantly outperforms the trust-related models, including the two baseline methods and the commute-time kernel alone (that is also based on social influence theory). In fact, if we evaluate the seven kernels in this approach individually, three of them, the impact distribution kernel ($K_{ID}$) and the two action kernels ($K_{ACT1}$, $K_{ACT2}$), would have their RMSE at the 1.3~1.4 level, which is already better than the two trust-based methods. The other four kernels ($K_{CT}$, $K_{COM}$, $K_{DEM}$, and $K_{CLA}$), on the other hand, have a very high RMSE, which indicates that applying SVR on them may not converge at all. A strength of our NLMKL method is its capability in combining information from different theories for a performance improvement. Of the two trust-based models, using multiple levels of friendships achieves better performance than using direct neighbors, which is consistent with previous studies' findings. Under the collaborative filtering paradigm, many models have slightly better performance than their SVR correspondents, since they do not have the convergence problem. However, they have less power to boost performance to a higher level as compared with our multi-kernel approach. By considering user bias, the CF model can avoid unpredictable users by providing a default rating. However, the overall prediction performance is not superior. It should be noted that the impact distribution kernel we designed works well in both the kernel-based paradigm and the collaborative filtering paradigm.

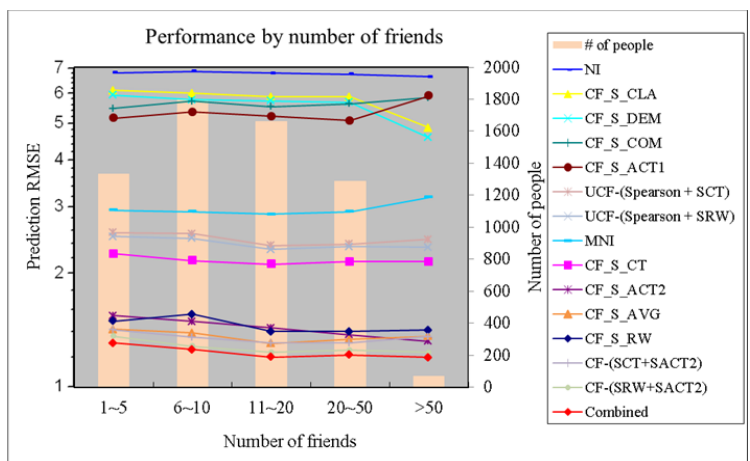

Fig. 4. Performance by Number of Friends

To further illustrate the effectiveness of our proposed approach, we investigate whether the number of friends may have an effect on prediction performance. In our dataset, the maximum number of friends is 81. We divide all users into five groups by the number of friends they had: 1-5, 6-10, 11-20, 20-50, >50. As we can see in Fig. 4, the first four groups generally have 1200 to 1800 users. Fig. 4 shows the RMSE according to the five groups of users. In general, most models show an increasing performance (i.e., decreasing RMSE) on users with more friends. Our proposed approach keeps this trend. Its RMSE decreases approximately 8% from people with less than 5 friends to people with more than 10 friends. This curve shows our model's effectiveness in taking advantage of information from a uses' friends in inferring user opinion.

Table 4 reports the weight η of all kernels in the NLMKL. In this combination process, we include an all-one matrix to capture first-order kernel components. In the combination, the weight for this kernel is 0.388, which indicates that first-order components, i.e., the seven original kernels, occupy a significant portion of the final combined kernel. The action kernels, which have a good performance individually, and the commute time kernel, which performs so-so individually, also occupy a large portion of the final kernel. This shows that NLMKL is able to take advantage of hidden power of kernels by interacting with other kernels even if the kernel does not perform well individually. The impact distribution kernel, which performs well, occupies a small but significant portion of the final kernel. It could also be part of the reason for the combined kernel's good performance. While the usefulness of each kernel cannot be fully assessed through these weights, they indicate that the kernels directed by contagion theory (impact distribution kernel and commute time kernel) and homophily theory (action kernels) can play important roles in a recommender system application.

Table 4. Weight for Each Kernel from NLMKL

| Kernel | ONES | $K_{ID}$ | $K_{CT}$ | $K_{COM}$ | $K_{DEM}$ | $K_{CLA}$ | $K_{ACT1}$ | $K_{ACT2}$ |
|---|---|---|---|---|---|---|---|---|
| Average η | 0.388 | 0.009 | 0.559 | 0.009 | 0.008 | 0.006 | 0.518 | 0.463 |

# 6. Discussion and Conclusions

In this paper, we propose a multi-theoretical approach that allows us to capture and combine information on different facets' of social networks for product recommendation. We have selected several kernels and developed an impact distribution kernel to reflect multiple social network theories. This effort bridges the parallel explorations in social science and data mining on the (social network-based) recommendation problem. In experiments on a real-world dataset, the multi-theoretical approach is significantly better than popular trusted-based models and the collaborative filtering approach. The performance of our proposed approach increased quickly as users' numbers of friends increased. Further analysis shows that kernels based on contagion theory and homophily theory contribute more to the formation of combined kernels, while other kernels also affect the performance of the approach.

In the future, we will explore how to extend this multi-theoretical framework. 1) We will continue to study the mapping between kernels and social network theories and explore kernels that better align with the insight of social network theories. 2) We will investigate advanced MKL approaches that fit the requirement of recommender systems. 3) We observe that the kernel-based method is more computationally expensive than collaborative filtering methods at the training stage. In general, we think it has a value for middle- and small-scale applications and for applications that do not have a strict resource limitation at the training stage. Nevertheless, we will investigate approaches that can improve the computational efficiency of our approach while keeping its good performance. Our ultimate goal is to build a unified and effective approach that is scalable to incorporate insights from different social theories for recommendation.

**Acknowledgements**

We appreciate the anonymous reviewers' valuable comments. We thank Mr. Xin Su's help on data collection. This work was partially supported by City University of Hong Kong SRG 7003008, 7002898 and 7002625. All opinions are those of the authors and do not necessarily reflect the views of the funding agencies.

**Xin Li** is an Assistant Professor in the Department of Information Systems at the City University of Hong Kong. He received the B. Eng. degree and the M. Eng. degree from Tsinghua University, and the Ph.D. degree in management information systems from the University of Arizona. His work has appeared in the Journal of MIS, Decision Support Systems, JASIST, Journal of Biomedical Informatics, Bioinformatics, among others. His current research interests include business intelligence & knowledge discovery, social network analysis, social media, and scientometric analysis.

**Mengyue Wang** is a PhD student in the Department of Information Systems at City University of Hong Kong. She received the M. Eng. Degree from University of Science of Technology of China. Her research interests include recommendation systems and electronic commerce. Her work has been published in the International Conference on Information Systems.

**Ting-Peng Liang** is the Director of Service Innovation and Electronic Commerce Research Center at the National Chengchi University in Taiwan. He is also a Fellow of the Association for Information Systems. He received his doctoral degree in Information Systems from the Wharton School of the University of Pennsylvania and had taught at the University of Illinois, Purdue University, Chinese University of Hong Kong, and City University of Hong Kong. His primary research interests include electronic commerce, intelligent systems, decision support systems, knowledge management, and strategic applications of information systems. His papers have appeared in journals such as Management Science, MIS Quarterly, Journal of MIS, Operations Research, Decision Support Systems, and Decision Sciences. He also serves on the editorial board of Decision Support Systems and several academic journals.